\documentclass[print,superscriptaddress,onecolumn]{revtex4-1}
\usepackage{amssymb}
\usepackage{amsmath}
\usepackage{graphicx}
\usepackage{color}
\usepackage{subfigure}
\usepackage{epstopdf}
\usepackage{geometry}
\begin{document}
\def\be{\begin{eqnarray}}
\def\ee{\end{eqnarray}}
\def\Tr{\mathrm{Tr}}

\title{Entanglement witnesses based on symmetric informationally complete measurements}
\author{Tao Li}
\affiliation{School of Mathematics and Statistics, Beijing Technology and Business University, Beijing 100048, China}
\author{Le-Min Lai}
\thanks{lailemin@outlook.com}
\affiliation{School of Mathematical Sciences,  Capital Normal University,  Beijing 100048,  China}
\author{Deng-Feng Liang}
\affiliation{School of Mathematics and Statistics, Beijing Technology and Business University, Beijing 100048, China}
\author{Shao-Ming Fei}
\thanks{feishm@cnu.edu.cn}
\affiliation{School of Mathematical Sciences,  Capital Normal University,  Beijing 100048,  China}
\affiliation{Max Planck Institute for Mathematics in the Sciences, Leipzig 04103, Germany}
\author{Zhi-Xi Wang}
\thanks{wangzhx@cnu.edu.cn}
\affiliation{School of Mathematical Sciences,  Capital Normal University,  Beijing 100048,  China}

\begin{abstract}
~\\
\section*{Abstract}
\baselineskip20pt
\noindent We study entanglement witness and present a construction of entanglement witnesses in terms of the symmetric informationally complete measurements (SIC-POVM). The capability of our witness is shown by some examples and it can be found this witness detects more entanglement than previous separability method given also by SIC-POVM. What's more, comparing with the dimension dependence of SIC-POVM, we also give the entanglement witnesses can be always constructed from general symmetric informationally complete measurements (GSIC-POVM).
\\~\\
\textbf{Keywords: Entanglement Witness, SIC-POVM}
\end{abstract}

\maketitle

\section{Introduction}
\baselineskip22pt
Quantum entanglement is an important resource that cannot be ignored in the field of quantum information processing\cite{1, 2, 3}. However, in practical applications, determining whether a given state is entangled is a difficult problem \cite{4, 5, 6}. It is essential to develop simple and effective entanglement detection methods, for example, positive partial transposition (PPT) criterion \cite{7,8,9}, realignment criterion \cite{10, 11, 12, 13, 14, 15, 16}, covariance matrix criterion \cite{17}, and correlation matrix criterion \cite{18, 19} are used to distinguish quantum entangled states from the separable ones.

Although for a given known quantum state there are many mathematical tools have been used in entanglement detection, problems related to the experimental implementation of entanglement detection for unknown quantum states have rarely yielded results \cite{20,21, 22, 23, 24}. By taking use of mutually unbiased bases (MUB) \cite{25}, the authors deal with the separability problem in two-qubit,  multipartite and continuous-variable quantum systems \cite{26}. These entanglement criteria are not only powerful, but also can be implemented experimentally. Furthermore, such idea can be generalized to the entanglement criterion based on mutually unbiased measurements (MUM) \cite{27,28,29}. Besides the MUB and MUM, symmetric informationally complete measurements (SIC-POVM) is another intriguing topic in quantum information theory 
and enormous work is completed in recent years. In order to avoid the open question that  whether SIC-POVM exists in arbitrary dimension, the authors introduced the general symmetric informationally complete measurements (GSIC-POVM) \cite{30}. Elements of GSIC-POVM are need not to be rank one and  GSIC-POVM exists in all finite dimensions. In fact, GSIC-POVM can be constructed from the generalized Gell-Mann matrices which showed by Gour and Kalev \cite{31}. Many of entanglement detection methods have been given for $d$-dimensional bipartite systems based on SIC-POVM and GSIC-POVM \cite{32,33,34,35,36,37,38,39}.

On the other hand, the entanglement witness (EW) is regarded as a useful approach to characterize the quantum entanglement \cite{40}. Detecting an unknown quantum state by EW requires less information than state tomography. It only concerns the mean values of some observables for a given quantum state. An entanglement witness $W$ is an hermitian operator which is block-positive but not positive. A bipartite state $\rho$ is separable if and only if $\Tr (\rho W)\geqslant0$ for all entanglement witnesses. It would be interesting to combine the entanglement witness with POVM like MUB and MUM. In Refs. \cite{41,42}, the authors constructed the entanglement witness based on MUB and MUM. In this paper, we study the entanglement witness by using the symmetric informationally complete measurements. Detailed examples are presented to illustrate the advantages of these entanglement witnesses. Furthermore, we also give the entanglement witnesses from the general symmetric informationally complete measurements.

\section{Entanglement witness based on SIC-POVM}
In this section, we construct a class of entanglement witnesses based on SIC-POVM through a positive map.
A POVM $\{P_{1}, P_{2}, \cdots, P_{d^2}\}$ with $d^2$ rank one operators acting on $\mathbb{C}^d$ is symmetric informationally complete, if every operator is of the form $P_{j}=\displaystyle \frac{1}{d}|\phi_{j}\rangle\langle \phi_{j}|$, $j=1, 2, \cdots, d^2$, satisfying
\begin{itemize}
  \item $\mathrm{Tr}(P_j^2)=\displaystyle\frac{1}{d^2}$,
  \item $\mathrm{Tr}(P_jP_k)=\displaystyle\frac{1}{d^2(d+1)},~~~~~   j\neq k$.
\end{itemize}

In \cite{32}, the author, based on index of coincidence, derived a number of inequalities satisfied by the SIC-POVM $\{P_{1}, P_{2}, \cdots, P_{d^2}\}$ on $\mathbb{C}^d$ and the density matrix $\rho$,
\be \label{J}
\sum\limits^{d^2}\limits_{i=1}p_i(\mathcal{P}|\rho)=\sum\limits^{d^2}\limits_{i=1}[\Tr(P_i\rho)]^2=\displaystyle \frac{\Tr(\rho)^2+1}{d(d+1)}.
\ee
If $\rho$ is a pure state, in particular, one has $\sum\limits^{d^2}\limits_{i=1}[\Tr(P_i\rho)]^2=\displaystyle \frac{2}{d(d+1)}$.

Now, we are going to give an entanglement witness based on SIC-POVM through constructing a map $\Phi$ which is positive if and only if $\Phi$ maps rank one projector into a ball $\mathfrak{B}$ which takes the maximally mixed state as center (i.e. $\rho \in \mathfrak{B}\Leftrightarrow \mathrm{Tr}(\rho^2)\leqslant \frac{1}{d-1}$)\cite{42+1}. According to the relevant theorem in \cite{2}, we then show our witness by following theorem:

\textit{\textbf{Theorem 1.}} Let $\{P_1,P_2,\cdots,P_{d^2}\}$ denote the SIC-POVM on $\mathbb{C}^d$ and $\mathcal{O}$ be an orthogonal rotation in $\mathbb{R}^{d^2}$ around the axis $\vec{n}=(1,1,\cdots,1)/d$. Then the following map
\be
\Phi X = \frac{1}{d}\mathbb{I}_d\mathrm{Tr} X -\frac{d+1}{d-1}\sum\limits_{k,l=1}^{d^2}\mathcal{O}_{kl} \mathrm{Tr}(XP_{l}- \frac{1}{d}\mathbb{I}_d\mathrm{Tr} XP_{l} )P_{k},
\ee
where the $\displaystyle \frac{1}{d}\mathbb{I}_d\mathrm{Tr} X$ denotes completely depolarizing channel and $P_{l},P_{k}$ are elements of $\{P_1,P_2,\cdots,P_{d^2}\}$ , is positive. Furthermore,
\be
W_{\Phi}=\displaystyle\frac{2}{d(d+1)}\mathbb{I}_d\otimes\mathbb{I}_d- \sum\limits_{k,l=1}^{d^2}\mathcal{O}_{kl} \overline{P}_{l} \otimes P_{k}
\ee
is an entanglement witness, where $\overline{P}_{l} $ denotes the conjugation of $P_{l}$ and $\mathcal{O}_{kl}$ are elements of orthogonal rotation $\mathcal{O}$.

\textit{Proof.} For any rank-1 projector $P = |\phi\rangle\langle \phi|$, we have
\begin{widetext}
\begin{eqnarray}
\mathrm{Tr}(\Phi P)^2 &=&\mathrm{Tr}\left[\frac{1}{d^2}\mathbb{I} -2\frac{d+1}{d-1}\frac{1}{d} \sum\limits_{k,l=1}^{d^2}\mathcal{O}_{kl} \mathrm{Tr}(PP_{l}-\frac{1}{d}\mathbb{I}_d\Tr P P_l )P_{k} \right.\nonumber\\  &&~~~~~\left.+(\frac{d+1}{d-1})^2 \sum\limits_{k,l,m,n=1}^{d^2}\mathcal{O}_{kl} \mathrm{Tr}(PP_{l}-\frac{1}{d}\mathbb{I}_d\Tr P P_l )P_{k} \mathcal{O}_{mn} \mathrm{Tr}(PP_{n}-\frac{1}{d}\mathbb{I}_d\Tr P P_n) P_{m}
\right]\nonumber\\
&=&\frac{1}{d}-2\frac{1}{d}\frac{d+1}{d-1} \sum\limits_{k,l=1}^{d^2}\mathcal{O}_{kl} \mathrm{Tr}(PP_{l}-\frac{1}{d}\mathbb{I}_d\Tr P P_l  )\Tr(P_{k})\nonumber\\
&&+(\frac{d+1}{d-1})^2  \sum\limits_{k,l,m,n=1}^{d^2}\mathcal{O}_{kl} \mathcal{O}_{mn} \mathrm{Tr}(PP_{l}-\frac{1}{d}\mathbb{I}_d\Tr P P_l  )\mathrm{Tr}(PP_{l}-\frac{1}{d}\mathbb{I}_d\Tr P P_n  )\Tr(P_{k} P_{m} )\nonumber\\
&=&\frac{1}{d}+(\frac{d+1}{d-1})^2  \left[ \sum\limits_{k=m=1}^{d^2}\sum\limits_{l,n=1}^{d^2}\mathcal{O}_{kl} \mathcal{O}_{kn} \mathrm{Tr}(PP_{l}-\frac{1}{d}\mathbb{I}_d\Tr P P_l  )\mathrm{Tr}(PP_{n}-\frac{1}{d}\mathbb{I}_d\Tr P P_n  )\Tr(P_{k} P_{k} ) \right.\nonumber\\  &&~~~~~~~~~~~~~~~~~~~~\left.+\sum\limits_{k\neq m=1}^{d^2}\sum\limits_{l,n=1}^{d^2}\mathcal{O}_{kl} \mathcal{O}_{mn} \mathrm{Tr}(PP_{l}-\frac{1}{d}\mathbb{I}_d\Tr P P_l )\mathrm{Tr}(PP_{n}-\frac{1}{d}\mathbb{I}_d\Tr P P_n )\Tr(P_{k} P_{m} )
\right]\nonumber\\
&=&\frac{1}{d}+(\frac{d+1}{d-1})^2 \left[ \sum\limits_{l,n=1}^{d^2}\delta_{ln}\mathrm{Tr}(PP_{l}-\frac{1}{d}\mathbb{I}_d\Tr P P_l )\mathrm{Tr}(PP_{n}-\frac{1}{d}\mathbb{I}_d\Tr P P_n )\frac{1}{d^2}\right.\nonumber\\  &&~~~~~~~~~~~~~~~~~~~~\left.+\sum\limits_{k\neq m=1}^{d^2}\sum\limits_{l,n=1}^{d^2}\mathcal{O}_{kl} \mathcal{O}_{mn} \mathrm{Tr}(PP_{l}-\frac{1}{d}\mathbb{I}_d\Tr P P_l )\mathrm{Tr}(PP_{n}-\frac{1}{d}\mathbb{I}_d\Tr P P_n )\frac{1}{d^2(d+1)}\right]\nonumber\\
&=&\frac{1}{d}+\frac{1}{d^2}(\frac{d+1}{d-1})^2\sum\limits_{l=1}^{d^2}[\Tr(PP_{l}-\frac{1}{d}\mathbb{I}_d\Tr P P_l)]^2, \nonumber
\end{eqnarray}
\end{widetext}
where the second equality is due to $\mathcal{O}\vec{n}=\pm \vec{n}$. Combining with Eq. (\ref{J}) we arrive at
$\mathrm{Tr}(\Phi P)^2<{1}/({d-1})$. Therefore, $\Phi$ is a positive map.
The corresponding entanglement witness
$W_{\Phi}$ can be obtained from $W_{\Phi}=\displaystyle\frac{d-1}{d+1}\sum\limits_{i,j=1}^{d}|i\rangle\langle j|\otimes \Phi |i\rangle\langle j|$ directly. $\square$

Now, let us consider some examples to show the validity of the entanglement witness $W_{\Phi}$ based on SIC-POVM.

\noindent\textit{Example 1.} We first consider the maximally entangled state $|\phi^{+}\rangle=\displaystyle\frac{1}{\sqrt{d}}\sum\limits_{i=1}\limits^{d}|ii\rangle$. Taking $\mathcal{O}=\mathbb{I}_d$, we have
$$
\begin{array}{rcl}
\mathrm{Tr}\left(W_{\Phi}(|\phi^{+}\rangle\right)&=&\displaystyle\mathrm{Tr}\left(\frac{2}{d^2(d+1)}\sum\limits_{i=1}^{d}\mathbb{I}_d|i\rangle\langle i|\otimes \mathbb{I}_d|i\rangle\langle i|-\frac{1}{d}\sum\limits_{l=1}^{d^2}\sum\limits_{i=1}^{d}\overline{P}_{l}|i\rangle\langle i|\otimes P_{l}|i\rangle\langle i|\right)\\[6mm]
&=&\displaystyle\frac{2}{d(d+1)}-\frac{1}{d}\\[4mm]
&=&\displaystyle\frac{-d+1}{d(d+1)}<0.
\end{array}
$$
Thus the entanglement witness detects all the maximally entangled states.

\noindent\textit{Example 2.} Also, theorem 1 allows us to detect noise levels of mixtures of a maximally entangled state with white noise:
\begin{eqnarray}
\rho_{\mathrm{iso}}=\alpha|\phi^{+}\rangle \langle \phi^{+}|+\frac{1-\alpha}{d^2}\mathbb{I},
\end{eqnarray}
where $|\phi^{+}\rangle=\displaystyle\frac{1}{\sqrt{d}}\sum\limits_{i=1}\limits^{d}|ii\rangle$, $0 \leqslant \alpha <1$. We also take $\mathcal{O}=\mathbb{I}_d$ and get
\be\label{13}
\Tr(W_{\Phi}\rho_{iso})&=&\Tr\left[(\frac{2}{d(d+1)}\mathbb{I}_d\otimes\mathbb{I}_d-\sum\limits_{l=1}^{d^2}\overline{P}_{l}\otimes P_{l})\rho_{iso}\right]\nonumber\\
&=&\frac{2}{d(d+1)}-\frac{(d-1)\alpha+1}{d^2}.
\ee
If $\alpha>\displaystyle\frac{1}{d+1}$, then $\Tr(W_{\Phi}\rho_{iso})<0$ and $\rho_{iso}$ must be entangled. That is to say $W_{\Phi}$ detects all the entanglement in isotropic states which coincides with the one given in \cite{43}. In other words, The separability of $\rho_{iso}$ detected by $W_{\Phi}$ indeed necessary and sufficient.

\noindent\textit{Example 3.} Now, we can consider a more general example that following state $\rho$ given by
\be
\tiny
\left(
\begin{array}{ccccccccc}
 0.18 & 0.19-0.015 i & 0.064+0.093 i & -0.21+0.047 i & 0.096-0.083 i & 0.037-0.072 i & -0.061-0.00074 i & 0.09-0.052 i & 0.098-0.042 i \\
 0.19+0.015 i & 0.21 & 0.062+0.11 i & -0.23+0.034 i & 0.11-0.083 i & 0.046-0.075 i & -0.066-0.0059 i & 0.1-0.049 i & 0.11-0.037 i \\
 0.064-0.093 i & 0.062-0.11 i & 0.072 & -0.052+0.13 i & -0.0093-0.081 i & -0.025-0.045 i & -0.023+0.032 i & 0.0052-0.067 i & 0.014-0.067 i \\
 -0.21-0.047 i & -0.23-0.034 i & -0.052-0.13 i & 0.27 & -0.14+0.075 i & -0.064+0.077 i & 0.073+0.017 i & -0.12+0.039 i & -0.13+0.024 i \\
 0.096+0.083 i & 0.11+0.083 i & -0.0093+0.081 i & -0.14-0.075 i & 0.092 & 0.054-0.022 i & -0.033-0.029 i & 0.074+0.014 i & 0.073+0.024 i \\
 0.037+0.072 i & 0.046+0.075 i & -0.025+0.045 i & -0.064-0.077 i & 0.054+0.022 i & 0.037 & -0.012-0.025 i & 0.04+0.026 i & 0.037+0.031 i \\
 -0.061+0.00074 i & -0.066+0.0059 i & -0.023-0.032 i & 0.073-0.017 i & -0.033+0.029 i & -0.012+0.025 i & 0.021 & -0.031+0.019 i & -0.034+0.015 i \\
 0.09+0.052 i & 0.1+0.049 i & 0.0052+0.067 i & -0.12-0.039 i & 0.074-0.014 i & 0.04-0.026 i & -0.031-0.019 i & 0.062 & 0.062+0.0077 i \\
 0.098+0.042 i & 0.11+0.037 i & 0.014+0.067 i & -0.13-0.024 i & 0.073-0.024 i & 0.037-0.031 i & -0.034-0.015 i & 0.062-0.0077 i & 0.064 \\
\end{array}
\right).\nonumber
\ee
 Taking the orthogonal rotation $\mathcal{O}$ and the SIC-POVM operators shown in Appendix, we can directly obtain the witness $W_{\Phi}$ given by
\be
\tiny
\left(
\begin{array}{ccccccccc}
 0.006 & -0.02-0.009 i & -0.009-0.002 i & 0.005+0.006 i & -0.08-0.002 i & 0.01-0.004 i & 0.01-0.005 i & -0.01-0.002 i & -0.08-0.002 i \\
 -0.02+0.009 i & 0.08 & 0.009-0.009 i & 0.007-0.001 i & -0.005-0.006 i & -0.01+0.01 i & -0.01+0.007 i & -0.01+0.006 i & -0.002+0.005 i \\
 -0.009+0.002 i & 0.009+0.009 i & 0.08 & 0.01-0.002 i & -0.01-0.01 i & 0.003-0.001 i & 0.003-0.003 i & -0.005-0.002 i & 0.0003-0.004 i \\
 0.005-0.006 i & 0.007+0.001 i & 0.01+0.002 i & 0.08 & 0.02+0.009 i & 0.009+0.002 i & -0.01-0.01 i & 0.008+0.01 i & -0.005-0.003 i \\
 -0.08+0.002 i & -0.005+0.006 i & -0.01+0.01 i & 0.02-0.009 i & 0.007 & -0.009+0.009 i & 0.01+0.008 i & 0.01+0.01 i & -0.08+0.001 i \\
 0.01+0.0004 i & -0.01-0.01 i & 0.003+0.001 i & 0.009-0.002 i & -0.009-0.009 i & 0.08 & -0.003-0.005 i & 0.002+0.007 i & -0.0003-0.003 i \\
 0.01+0.0005 i & -0.01-0.007 i & 0.003+0.003 i & -0.01+0.01 i & 0.01-0.008 i & -0.003+0.005 i & 0.08 & -0.005-0.004 i & -0.0001-0.005 i \\
 -0.01+0.002 i & -0.01-0.006 i & -0.005+0.002 i & 0.008-0.01 i & 0.01-0.01 i & 0.002-0.007 i & -0.005+0.004 i & 0.08 & 0.0001-0.001 i \\
 -0.08+0.002 i & -0.002-0.005 i & 0.003+0.004 i & -0.005+0.003 i & -0.08-0.001 i & -0.003+0.003 i & -0.001+0.005 i & 0.0001+0.001 i & 0.001
\end{array}
\right).\nonumber
\ee
Therefore, we have $\Tr(W_{\Phi}\rho)=-0.0152221<0$, namely, the state is entangled.
In Ref. \cite{44}, a separability criterion has been presented based on SIC-POVM:
for any separable state $\rho$,
$G(\mathcal{N}_{AB}|\rho)\leq\displaystyle \frac{2}{d(d+1)}$, where $G(\mathcal{N}_{AB}|\rho)=\sum\limits_{j=1}^{d^2}P(j,j)$, $P(j,j)$ is a joint probability.
For this example, one has $G(\mathcal{N}_{AB}|\rho)-\displaystyle \frac{1}{6}=-0.011341<0$, i.e., this criterion can not detect the entanglement of the state.
\begin{figure*}[htbp]
\centering
\subfigure[The surface under the $0$-plan represents the Bell-diagonal states with the entanglement detected by $W_{\Phi}$.]{
\begin{minipage}[b]{0.53\linewidth}
\includegraphics[width=1\linewidth]{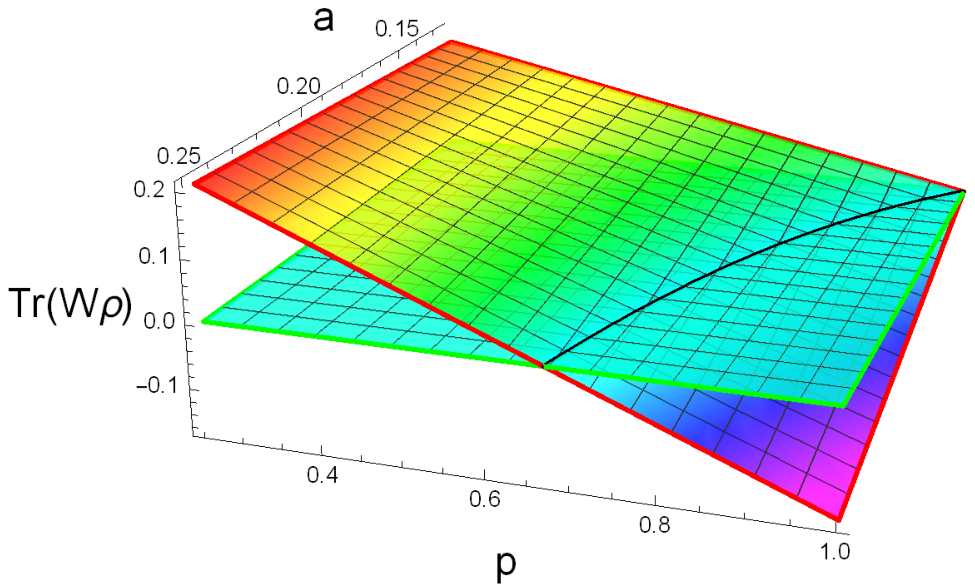}\vspace{-3pt}
\end{minipage}}~~~~~~~~~~
\subfigure[Entanglement detected by $W_{\Phi}$ for $a=0.2$ and $a=0.3$.]{
\begin{minipage}[b]{0.42\linewidth}
\includegraphics[width=1\linewidth]{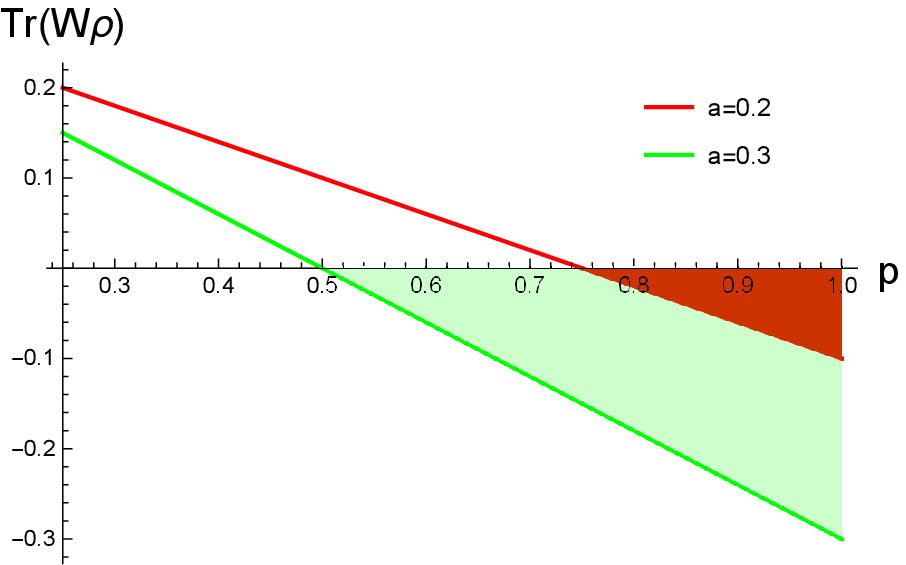}\vspace{10pt}
\end{minipage}}
\caption{Entanglement of the Bell-diagonal states detected by the witness $W_{\Phi}$ when $p>\displaystyle \frac{ad^2+1}{ad^2(d+1)}$. The witness detects more entanglement as $a$ increases.}
\end{figure*}

\section{Conclusion}
The separability problem via symmetric informationally complete measurements are studied in this paper and we show the quantum entanglement witnesses which are experimentally feasible for two-qudit. Some examples, such as maximally entangled states, white noise tolerance of maximally entangled states and a more general particular state are given to show the capability of such witnesses. Especially, for the given quantum state in the example, our witness is more powerful than the known entanglement criterion also based on SIC-POVM. Moreover, we have also presented an entanglement witness for arbitrary dimensional quantum systems based on the general symmetric informationally complete measurements. These results may highlight further investigations on experimentally feasible separability criteria.

\bigskip
{\sf Acknowledgments}~
This work is supported by the NSF of China under Grant No. 11675113, the Research Foundation for Youth Scholars of Beijing Technology and
Business University QNJJ2017-03, Scientific Research General Program of Beijing Municipal Commission of Education (Grant No.KM201810011009), NSF of Beijing under No. KZ201810028042, and Beijing Natural Science Foundation (Z190005), and Academy for Multidisciplinary Studies, Capital Normal University.

\newpage
\begin{center}
  Appendix
\end{center}
In example 3, we take the orthogonal rotation $\mathcal{O}$ and the SIC-POVM operators as follows:\\
the orthogonal rotation $\mathcal{O}$ is
\be
\left(
\begin{array}{ccccccccc}
 0.847 & 0.381 & 0.215 & -0.016 & 0.005 & -0.215 & -0.034 & -0.208 & 0.024 \\
 -0.131 & 0.825 & -0.321 & -0.013 & 0.004 & 0.321 & 0.015 & 0.311 & -0.011 \\
 -0.30 & 0.243 & 0.914 & -0.03 & 0.01 & 0.086 & -0.022 & 0.084 & 0.015 \\
 -0.015 & 0.038 & 0.022 & 0.998 & 0.001 & -0.022 & -0.003 & -0.021 & 0.002 \\
 0.005 & -0.013 & -0.007 & 0.001 & 1.00 & 0.007 & 0.001 & 0.007 & -0.001 \\
 0.30 & -0.243 & 0.086 & 0.03 & -0.01 & 0.914 & 0.022 & -0.084 & -0.015 \\
 0.011 & 0.016 & 0.028 & 0.001 & 0.00 & -0.028 & 0.999 & -0.027 & 0.001 \\
 0.291 & -0.235 & 0.084 & 0.03 & -0.01 & -0.084 & 0.021 & 0.919 & -0.015 \\
 -0.008 & -0.011 & -0.02 & -0.001 & 0.00 & 0.02 & 0.001 & 0.02
  & 0.999
\end{array}
\right)\nonumber
\ee
and the SIC-POVM operators are
\begin{eqnarray*}
P_1&=&\left(
\begin{array}{ccc}
 -0.066 & 0.017-0.017 i & 0.017-0.017 i \\
 0.017+0.017 i & 0.31 & 0.017-0.017 i \\
 0.017+0.017 i & 0.017+0.017 i & 0.091 \\
\end{array}
\right),~~~
P_2=\left(
\begin{array}{ccc}
 0.14 & -0.19-0.02 i & 0.017-0.017 i \\
 -0.19+0.02 i & 0.10 & 0.017-0.017 i \\
 0.017+0.017 i & 0.017+0.017 i & 0.091 \\
\end{array}
\right),
\\
P_3&=&\left(
\begin{array}{ccc}
 0.14 & 0.017-0.017 i & -0.19-0.02 i \\
 0.017+0.017 i & 0.10 & 0.017-0.017 i \\
 -0.19+0.02 i & 0.017+0.017 i & 0.091 \\
\end{array}
\right),~~~
P_4=\left(
\begin{array}{ccc}
 0.14 & 0.02+0.19 i & 0.017-0.017 i \\
 0.02-0.19 i & 0.10 & 0.017-0.017 i \\
 0.017+0.017 i & 0.017+0.017 i & 0.091 \\
\end{array}
\right),
\\
P_5&=&\left(
\begin{array}{ccc}
 0.020 & 0.017-0.017 i & 0.017-0.017 i \\
 0.017+0.017 i & -0.014 & 0.017-0.017 i \\
 0.017+0.017 i & 0.017+0.017 i & 0.33 \\
\end{array}
\right),~~~
P_6=\left(
\begin{array}{ccc}
 0.14 & 0.017-0.017 i & 0.017-0.017 i \\
 0.017+0.017 i & 0.10 & -0.19-0.02 i \\
 0.017+0.017 i & -0.19+0.02 i & 0.091 \\
\end{array}
\right),
\\
P_7&=&\left(
\begin{array}{ccc}
 0.14 & 0.017-0.017 i & 0.02+0.19 i \\
 0.017+0.017 i & 0.10 & 0.017-0.017 i \\
 0.02-0.19 i & 0.017+0.017 i & 0.091 \\
\end{array}
\right),~~~
P_8=\left(
\begin{array}{ccc}
 0.14 & 0.017-0.017 i & 0.017-0.017 i \\
 0.017+0.017 i & 0.10 & 0.02+0.19 i \\
 0.017+0.017 i & 0.02-0.19 i & 0.091 \\
\end{array}
\right),
\\
P_9&=&\left(
\begin{array}{ccc}
 0.22 & 0.068-0.068 i & 0.068-0.068 i \\
 0.068+0.068 i & 0.082 & 0.068-0.068 i \\
 0.068+0.068 i & 0.068+0.068 i & 0.033 \\
\end{array}
\right).
\end{eqnarray*}

\begin{thebibliography}{99}
\bibitem{1} M. A. Nielsen, I. L. Chuang,  \textit{Quantum Computation and
Quantum Information}. Cambridge University Press,  Cambridge,
2000.

\bibitem{2} R. Horodecki,  P. Horodecki,  M. Horodecki,  K. Horodecki,  Rev. Mod. Phys. \textbf{81},  865 (2009).

\bibitem{3} O. G\"uhne, G. T\'oth,  Phys. Rep. \textbf{474},  1 (2009).

\bibitem{4} L. Gurvits,  In Proc. of the 35th ACM Symp.  on Theory of Comp.  (ACM Press,  New York,  2003),  pp.  10-19.

\bibitem{5} S. Gharibian,  Quantum. Inf. Comput. \textbf{10},  343 (2010).

\bibitem{6} L. Gurvits,  J. Comput.  Syst. Sci. \textbf{69},  448 (2003).

\bibitem{7} A. Peres,  Phys. Rev. Lett.  \textbf{77},  1413 (1996).

\bibitem{8} M. Horodecki,  P. Horodecki,  R. Horodecki,  Phys. Lett. A  \textbf{223},  1 (1996).

\bibitem{9} P. Horodecki,  Phys. Lett.  A \textbf{232},  333 (1997).

\bibitem{10} O. Rudolph , Phys. Rev. A \textbf{67},  032312 (2003).

\bibitem{11} K. Chen,  L. A. Wu,   Quant. Inf. Comput. \textbf{3},  193 (2003).

\bibitem{12} M. Horodecki,  P. Horodecki,  R. Horodecki,  Open Syst. Inf. Dyn. \textbf{13},  103 (2006).

\bibitem{13} K. Chen,  L. A. Wu, Phys. Lett. A \textbf{306},  14 (2002).

\bibitem{14} K. Chen,  L. A. Wu,  Phys. Rev. A \textbf{69},  022312 (2004).

\bibitem{15} P. Wocjan,  M. Horodecki,  Open Syst. Inf. Dyn.  \textbf{12},  331 (2005).

\bibitem{16} S. Albeverio,  K. Chen,  S. M. Fei,  Phys. Rev. A \textbf{68},  062313 (2003).

\bibitem{17} O. Guhne, P. Hyllus, O. Gittsovich, J. Eisert, Phys. Rev. Lett. \textbf{99},  130504 (2007).

\bibitem{18} J. D. Vicente,  Quant. Inf. Comput. \textbf{7},  624 (2007).

\bibitem{19} J. D. Vicente,  J. Phys. A Math. Theor. \textbf{41},  065309 (2008).

\bibitem{20} M. Li,  J. Wang,  S. M. Fei,  X. Li-Jost,  Phys.  Rev.  A \textbf{89},  022325 (2014).

\bibitem{21} N. Gisin,  Phys. Lett. A \textbf{154},  201 (1991).

\bibitem{22} S. Yu,  J. W. Pan,  Z. B. Chen,  Y. D. Zhang,  Phys. Rev. Lett. \textbf{ 91},  217903 (2003).

\bibitem{23} M. Li,  S. M. Fei,  Phys. Rev. Lett. \textbf{104}, 240502 (2010).

\bibitem{24} M. J. Zhao,  T. Ma,  S. M. Fei,  Z. X. Wang,  Phys. Rev. A \textbf{83},  052120 (2011).

\bibitem{25} W. K. Wootters,  B. D. Fields,  Ann. Phys. (N. Y. ) \textbf{191},  363 (1989).

\bibitem{26} C. Spengler, M. Huber, S. Brierley, T. Adaktylos,  B. C. Hiesmayr, Phys.  Rev.  A \textbf{86},  022311 (2012).


\bibitem{27} B. Chen,  T. Ma,  S. M. Fei,  Phys. Rev.  A \textbf{89},  064302 (2014).

\bibitem{28} A. Kalev,  G. Gour,  New. J. Phys.  \textbf{16},  053038(2014).

\bibitem{29} L. Liu,  T. Gao, F. Yan,  arXiv: 1501. 01717 [quant-ph] (2015).

\bibitem{30} G. Gour,  A. Kalev,  J. Phys. A: Math.  Theor \textbf{47},  335302 (2014).

\bibitem{31} D. M. Appleby,  Opt.  Spectrosc. \textbf{103},  416 (2007).

\bibitem{32} A. E. Rastegin,  Eur.  Phys.  J.  D \textbf{67},  269 (2013).

\bibitem{33} B. Chen,  T. Li,  S. M. Fei,  Quantum. Inf. Process. \textbf{14}:2281-2290 (2014).

\bibitem{34} Y. Xi,  Z. J. Zheng,  C. J. Zhu,  Quantum. Inf. Process. \textbf{15},  5119 (2016).

\bibitem{35} J. W. Shang, A. Ali, H. J. Zhu , et al., Phys. Rev. A. \textbf{98},022309 (2018).

\bibitem{36}  L. Liu, T. Gao, F. L. Yan, Science China, \textbf{10},7-12 (2017).

\bibitem{37}  Y. Y. Lu, et al., Int. J. Theor. Phys, \textbf{57},208-218 (2018).

\bibitem{38} S. Q. Shen, M. Li, X. Li-Jost, et al.,Quantum. Inf. Process. \textbf{17},111 (2018).

\bibitem{39} L. M. Lai, T. Li, S. M. Fei, Z. X. Wang, Quantum. Inf. Process. \textbf{17},314 (2018).

\bibitem{40} B. M. Terhal, Phys. Lett. A \textbf{271}, 319 (2000).

\bibitem{41} D. Chru\'ci\'ski, G. Sarbicki, F. Wudarski, Phys. Rev. A \textbf{97}, 032318 (2018).

\bibitem{42} T. Li, L. M. Lai, S. M. Fei, Z. X. Wang, Int. J. Theor. Phys, \textbf{58}, 3973-3985 (2019).

\bibitem{42+1} I. Bengtsson, K. \.{Z}yczkowski, Geometry of quantum states.\textit{ An introduction to quantum entanglement}. Rinton Press, Incorporated, (2008).


\bibitem{43} R. A. Bertlmann, K. Durstberger, B. C. Hiesmayr, P. Krammer, Phys. Rev. A \textbf{72}, 052331 (2005).

\bibitem{44} A. J. Scott, Grassl,M. J. Math. Phys. \textbf{51}, 042203 (2010).

\bibitem{45}D. M. Appleby,  Opt. Spectrosc. \textbf{103}, 416 (2007)
\end{thebibliography}
\end{document}